\documentclass[11pt,twoside]{article} 
\usepackage{cspm-asp2006}
\usepackage{epsfig,graphicx,natbib,url}  
\usepackage{lscape} 
\begin{document}
\setcounter{page}{539} 

\markboth{Joshi et al.}{North--South Asymmetry} 
\title{North--South Asymmetry of Solar Activity \\ during Cycle 23} 
\author{B. Joshi$^{1}$, 
        P. Pant$^{1}$, 
        P. K. Manoharan$^2$ 
        and K. Pandey$^{3}$}
\affil{$^1$ARIES, Nainital, India\\
       $^2$Radio Astronomy Centre, TIFR, Ooty, India\\
       $^3$Department of Physics, Kumaun University, Nainital, India}

\begin{abstract} 
In this paper, we have made a statistical analysis of solar H$\alpha$
flares that occurred during the period 1996 to 2005 to investigate
their spatial distribution with respect to northern and southern
hemispheres of the Sun. The analysis includes a total of 21608 single
events. The study shows a significant N$-$S asymmetry which is
persistent with the evolution of the solar cycle. The flare activity
favors the northern hemisphere in general during the rising and
maximum phase of the solar cycle (i.e., in 1997, 1999, and 2000), while
the declining phase (i.e., from 2001 to 2005) shows a southern
dominance.  Further, the monthly N$-$S asymmetry index for flares,
sunspot numbers and sunspot areas suggests similar variations for
these phenomena with the progress of solar cycle.  We also find that
in terms of asymmetric behavior of solar flares, cycle 23 seems to act
quite differently from cycle 22 but comparably to cycle 21.
\end{abstract}

\section{Introduction}
The occurrence of various features of solar activity is not symmetric
considering their manifestation in the northern and southern
hemispheres of the Sun. This phenomena, known as North$-$South (N$-$S)
asymmetry, has been studied using several solar activity indices such
as flares, sunspot numbers, sunspot areas, prominences \& filaments,
magnetic flux, coronal intensity, etc. by various authors (Newton \&
Milson 1955; Howard 1974; Roy 1977; Vizoso \& Ballester 1987; Garcia
1990; Verma 1993; Ata\c{c} \& \"{O}zg\"{u}\c{c} 1996; Li et al.\ 1998;
Temmer et al.\ 2002; Joshi \& Joshi 2004; Knaack et al.\ 2004;
Braj\v{s}a et al.\ 2005; Temmer et al.\ 2006).  These studies reveal the
existence of a real N$-$S asymmetry that has bearing with the solar
dynamo mechanism (Ossendrijver 1996).

In this paper a detailed analysis of N$-$S asymmetry has been
performed with solar H$\alpha$ flares that occurred during solar cycle
23. The study covers almost full solar cycle 23 (1996$-$2005) and is
an extension of our earlier analysis (Joshi \& Pant 2005). In
Section~2, the pattern of flare occurrence is studied with respect to
heliographic latitudes and the significance of observed N$-$S
asymmetry is evaluated using binomial probability distribution.  The
study also throw light on the evolution of flare activity with the
progress of solar cycle. Further, the comparison of asymmetric
behavior of flare activity is made with sunspot numbers and areas
which is presented in Section 3.  The results of the study are
discussed in the final section.

\section{N$-$S Distribution and Asymmetry of Solar Flares}
\begin{figure}
\centering
\includegraphics[width=8cm]{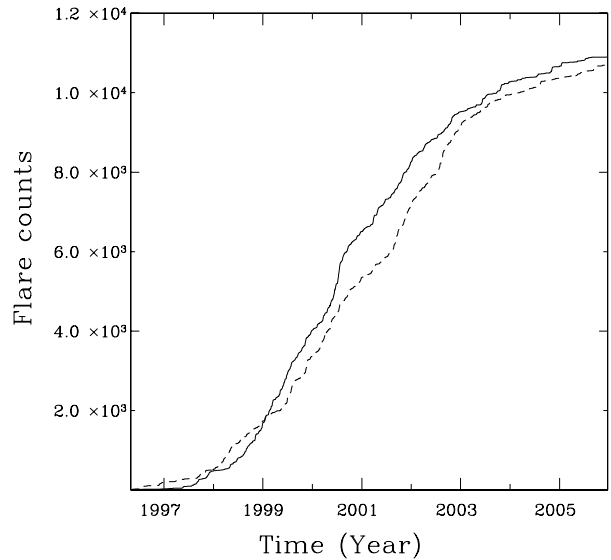}
\caption[]{ Cumulative counts of flares occurring in the northern (solid
line) and southern hemispheres (dashed line).}
\label{bj-fig:ns_cum_23}
\end{figure}

The data used in the present study have been collected from H$\alpha$
flare lists published in the SGD (Solar Geophysical Data) from 01 May
1996 to 31 December 2005, covering 10 years of solar cycle 23. Out of
21620 flares, 21608 events are selected for the study for which
complete information about the heliographic location (latitude and
longitude) was available.  To study the latitudinal distribution of
flares with solar cycle evolution, we compute the yearly flare counts
in $10^\circ$ latitudinal bands (within $\pm 50^\circ$ latitudes, flares
above $40^\circ$ latitudes being very small) in the northern and
southern hemispheres from 1996 to 2005 (Table 1). We find that the
yearly flare counts in the northern and southern hemispheres show an
asymmetry (column 8 of Table 1). To evaluate the statistical
significance of observed asymmetry we have used binomial probability
distribution (see Joshi \& Pant 2005 and references therein). The
cumulative binomial probability of getting more flares in one of the
hemispheres and consequently the dominant hemisphere is also given in
the table (see columns 9 and 10).  In Fig.~\ref{bj-fig:ns_cum_23},
N$-$S asymmetry has been presented by plotting the cumulative counts
of flares in the northern (solid line) and southern (dashed line)
hemispheres during cycle 23. The vertical distance between solid and
dashed lines is a measure of northern/southern excess up to that
time. To compare the N$-$S asymmetry observed in flares with sunspot
areas and numbers, we plot the monthly absolute asymmetry index
($\Delta = N_{\rm N} - N_{\rm S}$; see Ballester et al.\ 2005) for all these
phenomena in Fig.~\ref{bj-fig:month_asym}.

\begin{table}
\caption[]{Number of H$\alpha$ flares at different latitude bands in the
northern (N) and southern (S) hemispheres are tabulated for each year.
The binomial probability (Prob.) and the dominant hemisphere (DH) is
given for all the years as well as for all the latitudinal bands. A dash
($-$) specifies that the probability is not significant. Flares
occurring exactly at the equator have been excluded.}

\begin{center}
\scriptsize
\begin{tabular}{cccccccc|c|c}\hline
Years& \multicolumn{6} {c} {Number of flares}& &Prob.&DH\\ \cline{2-8}
   &$0$-$10^\circ$&$10$-$20^\circ$&$20$-$30^\circ$&$30$-$40^\circ$&$40$-$50^\circ$&$>50^\circ$&Total&&\\
\hline
1996\,N   &21   &2   &0   &1  &0  &0  &24  &1.107$\times$10$^{-33}$ &S  \\
~~~~~\, S &112  &53  &21  &2  &0  &0  &188 &                        &   \\
1997\,N   &39   &160 &230 &20 &0  &0  &449 &5.641$\times$10$^{-6}$  &N  \\
~~~~~\, S &5    &103 &199 &18 &1  &1  &327 &                        &   \\
1998\,N   &13   &635 &503 &54 &1  &1  &1207&0.428                   &$-$\\
~~~~~\, S &3    &366 &807 &31 &9  &0  &1216&                        &   \\
1999\,N   &169  &1240&811 &105&11 &0  &2336&7.017$\times$10$^{-30}$ &N  \\
~~~~~\, S &71   &874 &644 &38 &0  &0  &1627&                        &   \\
2000\,N   &463  &1327&633 &55 &1  &2  &2481&8.712$\times$10$^{-14}$ &N  \\
~~~~~\, S &248  &1288&373 &79 &1  &0  &1989&& \\
2001\,N   &503  &978 &262 &5  &1  &0  &1749&0.058                   &S  \\
~~~~~\, S &449  &1062&282 &42 &8  &0  &1843&                        &   \\
2002\,N   &250  &746 &255 &6  &0  &0  &1257&1.329$\times$10$^{-35}$ &S  \\
~~~~~\, S &675  &1006&272 &4  &0  &0  &1957&                        &   \\
2003\,N   &407  &315 &14  &3  &0  &0  &739 &0.069                   &S  \\
~~~~~\, S &222  &475 &83  &17 &0  &0  &797 &                        &   \\
2004\,N   &231  &159 &0   &0  &0  &0  &390 &0.123                   &$-$  \\
~~~~~\, S &165  &254 &4   &0  &0  &0  &423 &                   &  \\
2005\,N   &56   &176 &1   &0  &0  &1  &234 &9.054$\times$10$^{-6}$  &S  \\
~~~~~\, S &219  &117 &0   &0  &0  &0  &336 &                   &  \\

\hline
Total\,N  &2152 &5738&2709&249&14 &4  &10866 &0.134  &$-$\\
~~~~~\, S &2169 &5598&2685&231&19 &1  &10703 &  &\\
\hline
Prob.    &0.398&0.095&0.371&0.205&0.189&0.0625&0.134 &$-$&    \\
\cline{1-8}
DH      & $-$  &N   &$-$  &$-$ &$-$ &$-$ &&&    \\
\hline
\end{tabular}
\end{center}
\end{table}

\section{Results and Conclusions}

\begin{figure}
  \centering
  \includegraphics[width=8.5cm]{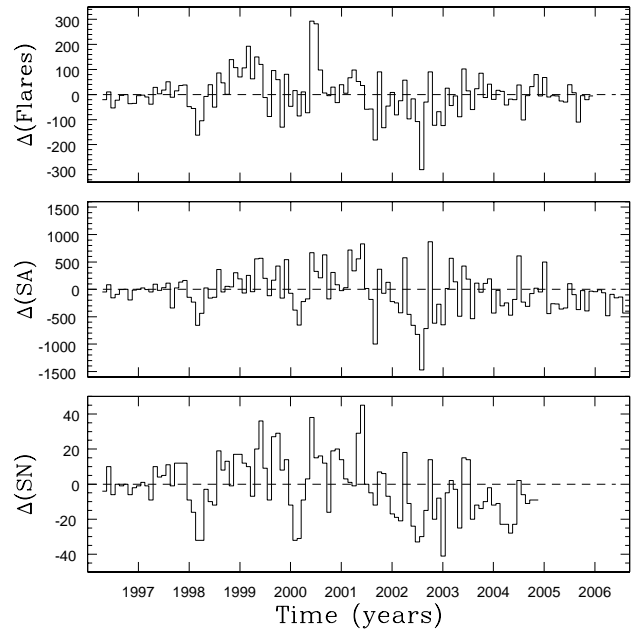}
  \caption {Monthly values of absolute asymmetry index ($\Delta =
N_{\rm N} - N_{\rm S}$) during solar cycle 23. The figure shows the plots of
asymmetry index for monthly flare counts, monthly mean sunspot areas
(SA) and monthly mean sunspot numbers (SN) (from top to bottom
panel).}
\label{bj-fig:month_asym}
\end{figure}

\leftmargini=3ex
\begin{enumerate} \itemsep=0ex \vspace*{-1ex}

\item
We find a significant N$-$S asymmetry in solar flare occurrence during
cycle 23 which is persistent during the course of the solar cycle. The
activity dominates the northern hemisphere in general during the
rising and maximum phase of the solar cycle (i.e., in 1997, 1999, and
2000), while the declining phase (i.e., from 2001 to 2005) shows
southern dominance.

\item
The cumulative flare counts show a southern excess in the rising phase
of the cycle (1997 to 1999), while after 1999 northern excess
prevails. The northern excess first increases till mid 2001 and then
decreases continuously till 2003. After 2003, we find a small but
constant northern excess (cf. Fig.~\ref{bj-fig:ns_cum_23}).  In
terms of asymmetric behavior of solar flares, cycle 23 seems to act
quite differently from the previous cycle (i.e.\ cycle 22) but is
comparable to cycle 21 (see Temmer et al.\ 2001 for N$-$S flare
asymmetry during cycles 21 and 22).  The different behavior of odd and
even numbered cycles may be interpreted as the two parts of the basic
22-year solar periodicity (\v{S}vestka, 1995).

\item
We find similar variations in N$-$S asymmetry of solar flares, sunspot
areas and sunspot numbers during solar cycle evolution
(cf. Fig.~\ref{bj-fig:month_asym}). However, the amplitude of N$-$S
asymmetry, determined by the absolute asymmetry index ($\Delta$), for
different months varies between various solar activity phenomena.

\end{enumerate}

\acknowledgements B. Joshi thanks the organizers for an excellent
conference that led to many fruitful discussions.

\end{document}